\documentclass[aps,twocolumn,showpacs,preprintnumbers,amsmath,amssymb,superscriptaddress,floatfix,nofootinbib]{revtex4}

\usepackage{graphicx}
\usepackage{epsfig}
\usepackage{epstopdf}
\usepackage{hyperref}
\usepackage{amsmath}
\usepackage{amsfonts}
\usepackage{amssymb}

\begin{document}

\title{The $\Xi^* \bar{K}$ and $\Omega \eta$ interaction within a chiral unitary approach}

\author{Siqi Xu}
\affiliation{College of Physics and Electronic Engineering,
Northwest Normal University, Lanzhou 730070, China}
\affiliation{Institute of modern physics, Chinese Academy of
Sciences, Lanzhou 730000, China}

\author{Ju-Jun Xie} \email{xiejujun@impcas.ac.cn}
\affiliation{Institute of modern physics, Chinese Academy of
Sciences, Lanzhou 730000, China} \affiliation{State Key Laboratory
of Theoretical Physics, Institute of Theoretical Physics, Chinese
Academy of Sciences, Beijing 100190, China}

\author{Xurong Chen}
\affiliation{Institute of modern physics, Chinese Academy of
Sciences, Lanzhou 730000, China}

\author{Duojie Jia}
\affiliation{College of Physics and Electronic Engineering,
Northwest Normal University, Lanzhou 730070, China}

\date{\today}

\begin{abstract}

In this work we study the interaction of the coupled channels
$\Omega \eta$ and $\Xi^* \bar{K}$ within the chiral unitary
approach. The systems under consideration have total isospins $0$,
strangeness $S = -3$, and spin $3/2$. We studied the $s$ wave
interaction which implies that the possible resonances generated in
the system can have spin-parity $J^P = 3/2^-$. The unitary
amplitudes in coupled channels develop poles that can be associated
with some known baryonic resonances. We find there is a dynamically
generated $3/2^-$ $\Omega$ state with mass around $1800$ MeV, which
is in agreement with the predictions of the five-quark model.

\end{abstract}


\maketitle

\section{Introduction}

Exploration of exotic hadrons that have more than three valence
quarks is an important issue in hadron physics. However, there is no
exotic hadron has been established so far, in contrast to the
existence of hundreds of ordinary hadrons. Very recently, the new
observation of the heavy hidden charm baryonic $P^+_c$
states~\cite{Aaij:2015tga} has challenged the conventional wisdom
that baryons are composed of three quarks from the naive quark
model. This observation has attracted a lot of attention from the
theoretical side. Various explanations of these states have been
proposed, such as molecules, muti-quark states, kinematic effects,
or mixtures of components of different nature. Nevertheless, up to
now none of them has been accepted unanimously.

In the case of light flavor baryons, only the nucleon and
$\Delta(1232)$ excited states have been abundantly studied both on
the theoretical and experimental sides, while for the cases of
hyperon excited states, the information of them is very scarce. For
example, there are only four $\Omega$ hyperon states compiled in the
"Review of Particle Physics" by the Particle Data Group
(PDG)~\cite{Agashe:2014kda}, namely the ground state $\Omega(1672)$,
and three excitations $\Omega(2250)$, $\Omega(2380)$, and
$\Omega(2470)$. Among these four states, only the ground state
$\Omega(1672)$ has been clarified to have spin-parity $J^P = 3/2^+$,
while the quantum numbers for the other three have not been
justified yet.

Although there has not been any further experimental evidence about
the $\Omega$ excited states sine the $1990$s, theorists are always
interested in the spectrum of the $\Omega$ hyperon, which has been
investigated within the traditional three quark models in
Refs.~\cite{Chao:1980em,Capstick:1986bm,Glozman:1995fu,Loring:2001ky},
the large $N_c$ expansion analysis in
Refs.~\cite{Carlson:2000zr,Schat:2001xr,Goity:2003ab,Matagne:2004pm,Matagne:2006zf},
the algebraic model in Ref.~\cite{Bijker:2000gq}, and the Skyrme
model in Ref.~\cite{Oh:2007cr}. Within these models, the predicted
mass of the lowest $3/2^-$ $\Omega$ excited states is around $2000$
MeV, which is always higher than the mass of the lowest $1/2^-$
$\Omega$ excited states. In Ref.~\cite{Wang:2008zzz}, the $\Xi
\bar{K}$ interaction was investigated within an extended chiral
$SU(3)$ quark model by solving a resonating group method equation.
It was shown that the $s$ wave $I =0$ $\Xi \bar{K}$ interaction is
attractive, and a $1/2^-$ $\Xi \bar{K}$ bound state with $3$ MeV
binding energy was predicted. Furthermore, in
Ref.~\cite{Wang:2007bf}, the $\Omega$ excited states in the $\Omega
\omega$ system with $J^P = 5/2^-$, $3/2^-$, and $1/2^-$ are
dynamically studied in both the chiral $SU(3)$ quark model and the
extended chiral $SU(3)$ quark model. The calculated results of that
reference show that the $\Omega \omega$ state has an attractive
interaction, and in the extended chiral $SU(3)$ quark model such
attraction can make for a $\Omega \omega$ quasi-bound state with
spin-parity $J^P = 3/2^-$ or $5/2^-$ and the binding energy of about
several MeV.

On the other hand, the spectrum of low-lying $\Omega$ states with
negative parity has been investigated by employing an extended
constituent quark model~\cite{Yuan:2012zs,An:2013zoa,An:2014lga},
within which the $\Omega$ states were cosidered as admixtures of
three- and five-quark components. It is shown that the mixing
between three- and five-quark components in $\Omega$ resonances with
spin-parity $J^P = 3/2^-$ is very strong, and the mixing decreases
the energy of the lowest $3/2^-$ state to be around $1785 \pm 25$
MeV, which is lower than that of the lowest $1/2^-$
state~\cite{An:2014lga}. Accordingly, five-quark components may be
more preferable in the wave function of those $\Omega$ excited
states.

The $\Xi^* \bar{K}$ and $\Omega \eta$ systems under consideration
have total isospin $I = 0$ and spin $J = 3/2$. If we considered only
the $s$ wave interaction of $\Xi^*$ and $\bar{K}$ or $\Omega$ and
$\eta$, then the possible resonances generated in the coupled
channels of $\Xi^* \bar{K}$ and $\Omega \eta$ can have only $J^P =
3/2^-$. Within the chiral unitary approach, the $s$ wave interaction
of the baryon decuplet with the octet of psedudoscalar mesons were
studied in Ref.~\cite{Sarkar:2004jh}. It was found that in the case
of strangeness $S = -3$ and isospin $I = 0$, there is a pole at
$(2141, -i38)$ MeV, which can be identified, by only the mass, with
the $\Omega(2250)$ resonance~\footnote{The other quantum numbers,
such as the spin and parity, of this state are unknown.} compiled in
the PDG. However, as discussed before, until now, the experimental
data for the $\Omega$ resonances is very poor. No $\Omega$ excited
states with negative parity have been observed yet. Further studies
about the $\Omega$ resonances are welcome.

In Ref.~\cite{Sarkar:2004jh}, it was claimed that the pole position
shifts with the value of the subtraction constant. Along this line,
in the present work, we re-study the $s$ wave interaction of the
coupled channels $\Xi^* \bar{K}$ and $\Omega \eta$ within the chiral
unitary approach. By adjusting the value of the subtraction
constant, we obtain a pole around $(1800,i0)$ MeV, which supports
the findings in Refs.~\cite{Yuan:2012zs,An:2013zoa,An:2014lga}. This
is very interesting that the energy of the lowest $3/2^-$ $\Omega$
state is lower than the energy of the lowest $1/2^-$ $\Omega$ state.

This paper is organized as follows. In next section, we discuss the
formalism and the main ingredients of the model. In
Sec.~\ref{sec:results}, we present our main results and, finally, a
short summary is given in Sec.~\ref{sec:summary}.

\section{Theoretical framework} \label{sec:formalism}

We begin with a brief discussion of the formalism of the chiral
unitary approach by reviewing the general procedure for calculating
the meson-baryon scattering amplitudes since more details can be
obtained from Ref.~\cite{Sarkar:2004jh}. In the chiral unitary
approach, from solving the Bethe-Salpeter equation, the scattering
matrix in coupled channels is given by~\cite{Sarkar:2004jh}
\begin{eqnarray}
T = [1 - VG]^{-1} V \, ,
\end{eqnarray}
where $V$ is the matrix for the transition potential between the
included channels and $G$, a diagonal matrix, is the loop function
for intermediate $\Xi^* {\bar K}$ and $\Omega \eta$ states, which is
defined as
\begin{eqnarray}
G &=& i \int \frac{d^4q}{(2\pi)^4} \frac{1}{q^2 - m^2 + i \epsilon}
\frac{2M}{(P-q)^2 - M^2 + i \epsilon} ,
\end{eqnarray}
where $m$ and $M$ are the masses of the $\bar{K}$ or $\eta$ meson
and the $\Xi^*$ or $\Omega$ baryon. In the above equation, $P$ is
the total incident momentum of the external meson-baryon system.

We study only the $s$ wave interaction, for which, the transition
potential for channel $i$ to $j$ reads~\cite{Sarkar:2004jh},
\begin{eqnarray}
V_{ij}&=& -C_{ij} \frac{1}{4f^2}(k^0+k^{\prime 0}),
\label{eq:potention}
\end{eqnarray}
with $f = 93$ MeV the pion decay constant. The $k^0$ and $k^{\prime
0}$ are the energy of the incoming and outgoing meson, respectively.
The transition coefficients $C_{ij}$ are symmetric with respect to
the indices, and also isospin-dependent. By naming the channels, $1$
for $\Xi^* \bar{K}$ and $2$ for $\Omega \eta$, the coefficients
$C_{ij}$ for the case of isospin $I = 0$ are~\cite{Sarkar:2004jh}
\begin{eqnarray}
C_{11} = 0,~~~~C_{12} = C_{21} = 3,~~~~C_{22} = 0.
\end{eqnarray}

From $C_{12} = C_{21} = 3$ and Eq.~\eqref{eq:potention}, we find an
attractive interaction between $\Xi^* \bar{K}$ and $\Omega \eta$
channels, for which we can expect bound states or resonances in the
coupled channels of $\Xi^* \bar{K}$ and $\Omega \eta$.

The loop function $G$ can be regularized either with a cutoff
prescription or with dimensional regularization in terms of a
subtraction constant. Here we make use of the dimensional
regularization scheme. The expression for $G$ is
then~\cite{Sarkar:2004jh,Ramos:2002xh}
\begin{eqnarray}
G_l &=& \frac{2M_l}{16\pi^2}\{a_l(\mu) + {\rm ln} \frac{M^2_l}{\mu^2} + \frac{m_l^2-M_l^2+s}{2s} {\rm ln} \frac{m_l^2}{M_l^2} \nonumber \\
&& + \frac{q_l}{\sqrt{s}}[{\rm ln} (s-(M_l^2-m_l^2) + 2q_l\sqrt{s}) \nonumber \\
&& + {\rm ln} (s+(M_l^2-m_l^2) + 2q_l\sqrt{s}) \nonumber \\
&& - {\rm ln}(-s+(M_l^2-m_l^2)+2q_l\sqrt{s}) \nonumber \\
&& - {\rm ln}(-s-(M_l^2-m_l^2)+2q_l\sqrt{s})]\} ,
\label{eq:gfunction}
\end{eqnarray}
with $\mu = 700$ MeV the scale of the dimensional regularization as
used in Ref.~\cite{Sarkar:2004jh}. Changes in the scale are
reabsorbed in the subtraction constant $a(\mu)$ through $a(\mu') -
a(\mu) = {\rm ln} \frac{\mu'^{2}}{\mu^2}$ so that the amplitude $T$
is scale independent. In Eq.(~\ref{eq:gfunction}), $q_l$ denotes the
three-momentum of meson or baryon in the center of mass frame, which
is given by:
\begin{eqnarray}
q_l = \frac{\lambda^{\frac{1}{2}}(s,m_l^2,M_l^2)}{2\sqrt{s}},
\end{eqnarray}
with $\lambda(x,y,z) = x^2 + y^2 + z^2 - 2xy - 2xz - 2yz$ being the
triangular function and $m_l$ and $M_l$ are the masses of the mesons
and baryons, respectively. In addition, we take $M_{\Xi^*} = 1533.4$
MeV, $m_{\bar{K}} = 495.6$ MeV, $M_{\Omega} = 1672.5$ MeV, and
$m_{\eta} = 547.9$ MeV.

The dynamically generated baryon states appear as poles of the
scattering amplitudes on the complex energy $\sqrt{s}$ plane. The
poles that are found on the second Riemann sheet are identified with
resonances. The mass and the width of the state can be found from
the position of the pole on the complex energy plane. In the second
Riemann sheet, the loop function $G$ in Eq.~\eqref{eq:gfunction}
should be changed, when ${\rm Re}(\sqrt{s})$ is above the $\Xi^*
\bar{K}$ ($2029$ MeV) or $\Omega \eta$ ($2220$ MeV) mass threshold,
with
\begin{eqnarray}
G_l^{II} = G_l + 2i \frac{q_l}{\sqrt{s}} \frac{M_l}{4\pi}, ~~ {\rm
with}~~{\rm Im}(q_l) > 0.
\end{eqnarray}

We have only two coupled channels, $\Xi^* \bar{K}$ and $\Omega
\eta$, and the transition potentials $V_{11} = V_{22} = 0$, then, we
search for the pole by looking for zero of the determinant of
$|1-VG|$
\begin{eqnarray}
{\rm det}|1-VG| = 1 - V^2_{12} G_{11} G_{22} = 0 ,
\end{eqnarray}
where $G_{11}$ and $G_{22}$ are the $G$ functions for $\Xi^*
\bar{K}$ and $\Omega \eta$ channels, respectively. In addition, the
scattering amplitudes of $T_{\Xi^* \bar{K} \to \Xi^* \bar{K}}$ and
$T_{\Omega \eta \to \Omega \eta}$ are obtained as,
\begin{eqnarray}
T_{\Xi^* \bar{K} \to \Xi^* \bar{K}} & = & \frac{V^2_{12}G_{22}}{1 -
V^2_{12} G_{11} G_{22}}, \\
T_{\Omega \eta \to \Omega \eta} & = & \frac{V^2_{12}G_{11 }}{1 -
V^2_{12} G_{11} G_{22}}.
\end{eqnarray}

Next, we determine the couplings of the resonance to different
channels, $\Xi^* \bar{K}$ and $\Omega \eta$ in the present case.
Close to the pole at $z_R$, the scattering amplitude behaves as
\begin{equation}
T_{ij} = \frac{g_ig_j}{\sqrt{s} - z_{\rm R}},    \label{eq:gij}
\end{equation}
where $g_i$ is the coupling of the state to the $i$-channel. We then
evaluate the residues of $T_{ij}$ to get the complex valued
couplings $g_i$.

\section{Numerical Results} \label{sec:results}

To evaluate the value of the scattering amplitudes $T$ we have to
fix the subtraction constants $a_l(\mu)$. Generally, $a(\mu)$ for
different channels is different, they should be determined by
fitting relevant experimental data. In the present work, we choose
the value of $a(\mu)$~\footnote{We take the same values for $\Xi^*
\bar{K}$ and $\Omega \eta$ channels.} to get the $\Omega$ state at
$1785$ MeV as corresponding to the estimated mass of $3/2^-$
$\Omega$ resonance in Ref.~\cite{An:2014lga}.

\begin{figure}[htbp]
\begin{center}
\includegraphics[scale=0.4]{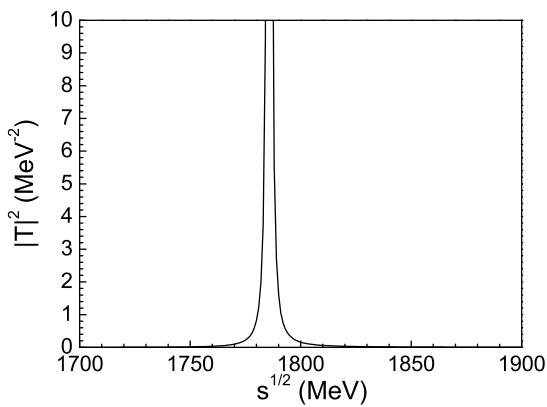}
\caption{Modulus squared $|T|^2$ of the $\Omega \eta \to \Omega
\eta$ transition. \label{Fig:tsquare}}
\end{center}
\end{figure}

With $a(\mu) = -3.4$, we obtain a pole of the $T$ matrix at $z_R =
(1785.7,-i0)$ MeV in the complex plane. The corresponding results of
$|T|^2$ as a function of $\sqrt{s}$ for $\Omega \eta \to \Omega
\eta$ transition is shown in Fig.~\ref{Fig:tsquare}, where there is
a clear peak around $1800$ MeV, which can be identified with the
$3/2^-$ $\Omega$ state that was predicted in Ref.~\cite{An:2014lga}.
Since the mass of this state is lower than the mass threshold of
$\Xi^* \bar{K}$ and $\Omega \eta$, it is a bound state of $\Xi^*
\bar{K}$ and $\Omega \eta$. Besides, because we do not include other
lower mass threshold decay channels, the obtained total decay width
of the $\Omega^*$ state is zero. If we take $a(\mu) = -2.0$, we can
also obtain a pole at $z_R = (2142.6,-i38.4)$ MeV as in
Ref.~\cite{Sarkar:2004jh}.

\begin{figure}[htbp]
\begin{center}
\includegraphics[scale=0.4]{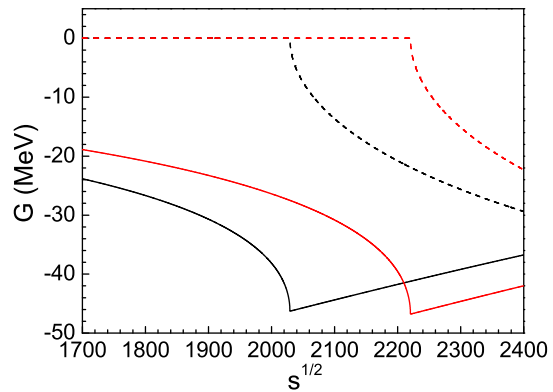}
\caption{(Color online) Real (solid lines) and imaginary (dashed
lines) parts of the loop function $G$ as a function of the total
scattering energy for the cases of $\Xi^* \bar{K}$ (black lines) and
$\Omega \eta$ (red lines). \label{Fig:gfunction}}
\end{center}
\end{figure}

In Fig.~\ref{Fig:gfunction}, we show the real and imaginary parts of
the loop function $G$ as a function of the total scattering energy.
The solid and dashed lines stand the results for the case of the
$\Xi^* \bar{K}$, while the red-solid and red-dashed lines represent
the case of the $\Omega \eta$. The results are obtained with $a(\mu)
= -3.4$.

The couplings, of the $3/2^-$ $\Omega$ state to $\Xi^* \bar{K}$
($g_{11}$) and $\Omega \eta$ ($g_{22}$) channels are also evaluated
around the pole [$z_R = (1785.7,-i0)$ MeV] using Eq.~\eqref{eq:gij},
which gives,
\begin{eqnarray}
g_{11} = 2.04, ~~~~ g_{22} = 2.31.
\end{eqnarray}
One can see that the $3/2^-$ $\Omega (1800)$ state has similar
couplings to $\Xi^* \bar{K}$ and $\Omega \eta$ channels.

\begin{figure}[htbp]
\begin{center}
\includegraphics[scale=0.4]{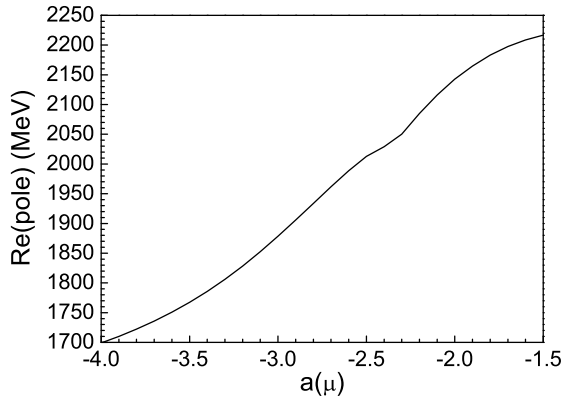}
\caption{Results of varying $a(\mu)$ over the $3/2^-$ $\Omega$
resonance mass. \label{Fig:realpolevsamu}}
\end{center}
\end{figure}

On the other hand, the true free parameter in the model is the
$a(\mu)$ subtraction constant in the loop functions, all other
parameters are meson masses or meson decay constants which are, in
principle, fixed by experiment. In view of this we shall vary the
value of $a(\mu)$ in the calculation to study the effect of $a(\mu)$
over the pole position of the $3/2^-$ $\Omega$ state. In
Fig.~\ref{Fig:realpolevsamu} one can see the effect of varying
$a(\mu)$ over the pole position of the $3/2^-$ $\Omega$ state. The
pole position moves from $1699$ MeV to $2217$ MeV with the parameter
$a(\mu)$ in the range of $-4.0 \leq a(\mu) \leq -1.5$. It is not
convenient for us to compare our results to the experimental data,
because the data are very poor~\cite{Agashe:2014kda}. While
comparing to predictions, $M_{\Omega^*} = 1785 \pm 25$ MeV, of
Ref.~\cite{An:2014lga}, we can get $a(\mu) = -3.4 \pm 0.1$. This
value is in line but a bit far from the natural size value of $-2$
that was used in Ref.~\cite{Sarkar:2004jh}, where a global fit to
the baryonic resonances from baryon decuplet and meson octet
interaction was conducted.

The results obtained here partly support the findings in
Refs.~\cite{Yuan:2012zs,An:2013zoa,An:2014lga} that the meson-baryon
components (or five-quark configuration) is more preferable in the
$3/2^-$ $\Omega$ excited states. This could be tested by future
experiments, as pointed in Ref.~\cite{An:2014lga}, the BESIII
experiment.

\section{Summary} \label{sec:summary}

In this work, within the chiral unitary approach, we have chosen the
$\Xi^* \bar{K}$ and $\Omega \eta$ systems as coupled channels to
investigate the dynamical generation of baryon excited states. The
systems under consideration have total isospins $0$, strangeness $S
= -3$, and spin $3/2$. We studied the $s$ wave interaction which
implies that the possible resonances generated in the system can
have spin-parity $J^P = 3/2^-$. The formalism consists of solving
Bethe-Salpeter equations. In the isospin $I = 0$ sector, by
adjusting the subtraction constant $a(\mu) = -3.4$, we find a bound
$\Omega$ excited state with mass around $1800$ MeV. This state can
be identified with the predicted $\Omega$ resonance with mass $M =
1785 \pm 25$ MeV in Ref.~\cite{An:2014lga}. It is shown that the
mass of this lowest $3/2^-$ $\Omega$ state is lower than the mass of
the lowest $1/2^-$ $\Omega$ state. Furthermore, there should be more
five-quark components in the wave function of the $3/2^-$ $\Omega$
state.

Finally, we would like to address that the value of $a(\mu) = -3.4$
is a bit far from the natural size value of $-2$, which was used in
Ref.~\cite{Sarkar:2004jh}. However, the experimental data of the
$\Omega^*$ states is so poor, hence, the value of $a(\mu)$ is still
open. It is expected that the future experiments about the
$\Omega^*$ states can provide more constraints on the value of
$a(\mu)$.

\section*{Acknowledgments}

We would like to thank Xu Cao for the useful discussions and
suggestions. This work is partially supported by the National Basic
Research Program (973 Program Grant No. 2014CB845406), and the
National Natural Science Foundation of China under Grant Nos.
11475227 and 11265014. It is also supported by the Open Project
Program of State Key Laboratory of Theoretical Physics, Institute of
Theoretical Physics, Chinese Academy of Sciences, China
(No.Y5KF151CJ1).

\end{document}